\documentclass[12pt]{article}
\usepackage{amssymb}
\usepackage{graphicx}
\usepackage[cp1251]{inputenc}
\usepackage{rotating}

\begin{document}
 \noindent {\footnotesize\it Astronomy Letters, 2023, Vol. 49, No. 6, pp. 320--330}
 \newcommand{\dif}{\textrm{d}}

 \noindent
 \begin{tabular}{llllllllllllllllllllllllllllllllllllllllllllll}
 & & & & & & & & & & & & & & & & & & & & & & & & & & & & & & & & & & & & & &\\\hline\hline
 \end{tabular}

  \vskip 0.5cm
  \bigskip
 \bigskip
 \centerline{\large\bf Determination of the Spiral Pattern Speed in the Milky Way}
 \centerline{\large\bf from Young Open Star Clusters}

 \bigskip
 \bigskip
  \centerline { 
   V. V. Bobylev\footnote [1]{vbobylev@gaoran.ru},  A. T. Bajkova}
 \bigskip
 \centerline{\small\it Pulkovo Astronomical Observatory, Russian Academy of Sciences, St. Petersburg, 196140 Russia}
 \bigskip
 \bigskip
{We have estimated the spiral pattern speed in the Galaxy $\Omega_p$ from a large sample of young open star clusters (OSCs). For this purpose, we have used 2494 OSCs younger than 50 Myr. Their mean proper motions, line-of-sight velocities, and distances were calculated by Hunt and Reffert (2023) based on data from the Gaia~DR3 catalogue. Three methods have been applied to estimate $\Omega_p$. They all are based on the linear Lin--Shu spiral density wave theory. We have obtained an estimate of $\Omega_p=24.26\pm0.52$~km s$^{-1}$ kpc$^{-1}$ by the first method, which is most reliable in our view, using the velocity perturbations $f_R$ and $f_\theta$ found through a spectral analysis of the radial, $V_R$, and residual rotation, $\Delta V_{\rm circ},$ velocities. Using the second method, we have found the velocity perturbations $f_R$ and $f_\theta$ by solving the basic kinematic equations together with the Galactic rotation parameters and obtained an estimate of
 $\Omega_p= 23.45\pm0.53$~km s$^{-1}$ kpc$^{-1}$. We have found 
 $\Omega_p= 28.9\pm2.8$~km s$^{-1}$ kpc$^{-1}$ by the third method based on an analysis of the position angles of OSCs at their birth time $\theta_{\rm birth}$.
 }

\bigskip
\section*{INTRODUCTION}
At present, there is great interest in studying the spiral structure of the Galaxy. This has become possible owing to the appearance of highly accurate
kinematic data on such spiral structure tracers as
masers, OB stars, or Cepheids. For example, the
spiral pattern speed was estimated in Bobylev and Bajkova (2023) from these objects.

Of course, the list of Galactic spiral structure tracers is wider. In particular, open star clusters (OSCs) are of great importance for studying the structure and kinematics of the Galaxy. Data on OSCs are used to estimate the parameters of the
Galactic rotation curve (Glushkova et al. 1998;
Zabolotskikh et al. 2002; Piskunov et al. 2006; Loktin
and Popova 2019; Popova 2023), the geometric
and kinematic characteristics of the Galactic spiral
density wave (Amaral and L\'epine 1997; Popova
and Loktin 2005; Naoz and Shaviv 2007; Bobylev et al. 2008; L\'epine et al. 2008; Junqueira et al. 2015; Camargo et al. 2015; Bobylev and Bajkova 2019; Cantat-Gaudin et al. 2020) as well as other structural and kinematic properties of OSCs (Kuhn et al. 2019; Tarricq et al. 2021; Monteiro et al. 2021).

The catalogues created as a result of the Gaia space project (Prusti et al. 2016) are an important source of extensive data on the trigonometric parallaxes, proper motions, and line-of-sight velocities of stars. In the Gaia EDR3 (Gaia Early Data Release~3) version (Brown et al. 2021) the trigonometric parallaxes
for $\sim$500 million stars were measured with errors less than 0.2~milliarcseconds (mas). The proper motions of about half of the stars in the catalogue were measured with a relative error less than 10\% (the error in the magnitude of the proper motion vector). In Gaia~EDR3 the line-of-sight velocities were copied from the previous version of the catalogue, Gaia~DR2 (Brown et al. 2018). In the last published version of the GaiaDR3 catalogue (Vallenari et al. 2022) the line-of-sight velocities of stars were improved significantly, while the parallaxes and proper motions of stars were simply copied from Gaia~EDR3.

The catalogues containing kinematic characteristics of OSCs are rapidly updated. For example, the catalogues by Dias et al. (2001, 2006, 2021) are well known. The last catalogue from this series presents 1743 OSCs whose fundamental parameters were determined from Gaia~DR2 data. These include estimates of the mean distances, ages, and proper motion components. The mean line-of-sight velocities were calculated for 831 clusters.

The catalogue by Cantat-Gaudin et al. (2020)
contains data on 2017 OSCs; 234 128 member stars
of these clusters turned out to be covered. The mean
OSC parameters were determined from GaiaDR2
data. An artificial neural network was used to estimate
the OSC extinction, distance modulus, and age.
In the opinion of these authors, reliable results were
obtained for 1867 OSCs.

The catalogue by Tarricq et al. (2021) gives the mean line-of-sight velocities for 1382 OSCs from the list by Cantat-Gaudin et al. (2020). The line-of-sight
velocities of stars were taken from various sources.
The weighted mean line-of-sight velocities for 38\%
of these OSCs may be deemed highly reliable, since they were calculated using more than three stars; the errors of such means are less than 3~km s$^{-1}$.

Hao et al. (2021) gave the mean proper motions and mean parallaxes of 3794 OSCs calculated from Gaia~EDR3 data. The age estimates were collected by
these authors from various sources. Hao et al. (2022) described another 704 previously unknown OSCs found from Gaia~EDR3 data. Thus, the catalogues of these authors form one of the most extensive kinematic databases of Galactic OSCs.

Note, finally, the paper by Joshi and Malhotra (2022), where more than 6000 clusters are presented. Information about the distances and ages from Gaia data is available for 4378 OSCs of them. It is based on such clusters that they analyzed their
spatial distribution and studied their kinematics. In particular, these obtained a new estimate of the spiral pattern speed in the Galaxy 
$\Omega_p=26.5\pm1.5$~km s$^{-1}$ kpc$^{-1}$ and the corotation radius 
$R_{\rm cor}=(1.08^{+0.06}_{-0.05})R_0.$ For this purpose, the position angles of OSCs at their birth time were analyzed.

The goal of this paper is to analyze the kinematics of the young OSCs presented in Hunt and Reffert (2023) that were identified by these authors already from Gaia~DR3 data.

 \section{DATA}
Hunt and Reffert (2023) presented 7200 OSCs that were identified from Gaia DR3 data. The popular HDBSCAN cluster analysis algorithm was used to identify the cluster members. New OSC candidates are 2420 from the total number of clusters found,
4780 OSCs are known from the literature, including
134 globular clusters. The more rigorous section of
the catalogue contains 4114 highly reliable OSCs, 749 of which are new. Such parameters as the age, lifetime, and distance were determined for all of the OSCs in the catalogue. A high percentage of OSCs in the catalogue have an estimate of the mean line-ofsight velocity from Gaia~DR3 data.

In this paper, to analyze the kinematics of the Galaxy and the spiral density wave, we selected OSCs younger than 50 Myr. At the same time, we used only OSCs marked by ``o'', i.e., moving groups and globular clusters were excluded. We did not use
any OSCs lying in the inner Galaxy with $R<4$~kpc, where there is a strong influence of the central bar. A total of 2494 clusters with a mean age of 21.1~Myr
were selected. The mean line-of-sight velocities were
calculated for 1722 OSCs of then. We divided this
sample into two approximately equal parts in age. The first included 1332 OSCs younger than 20~Myr with a mean age of 9.6 Myr. The second included 1162 OSCs with ages from the interval 20--50 Myr with a mean age of 34.3 Myr.

\begin{figure}[t]
{ \begin{center}
  \includegraphics[width=0.95\textwidth]{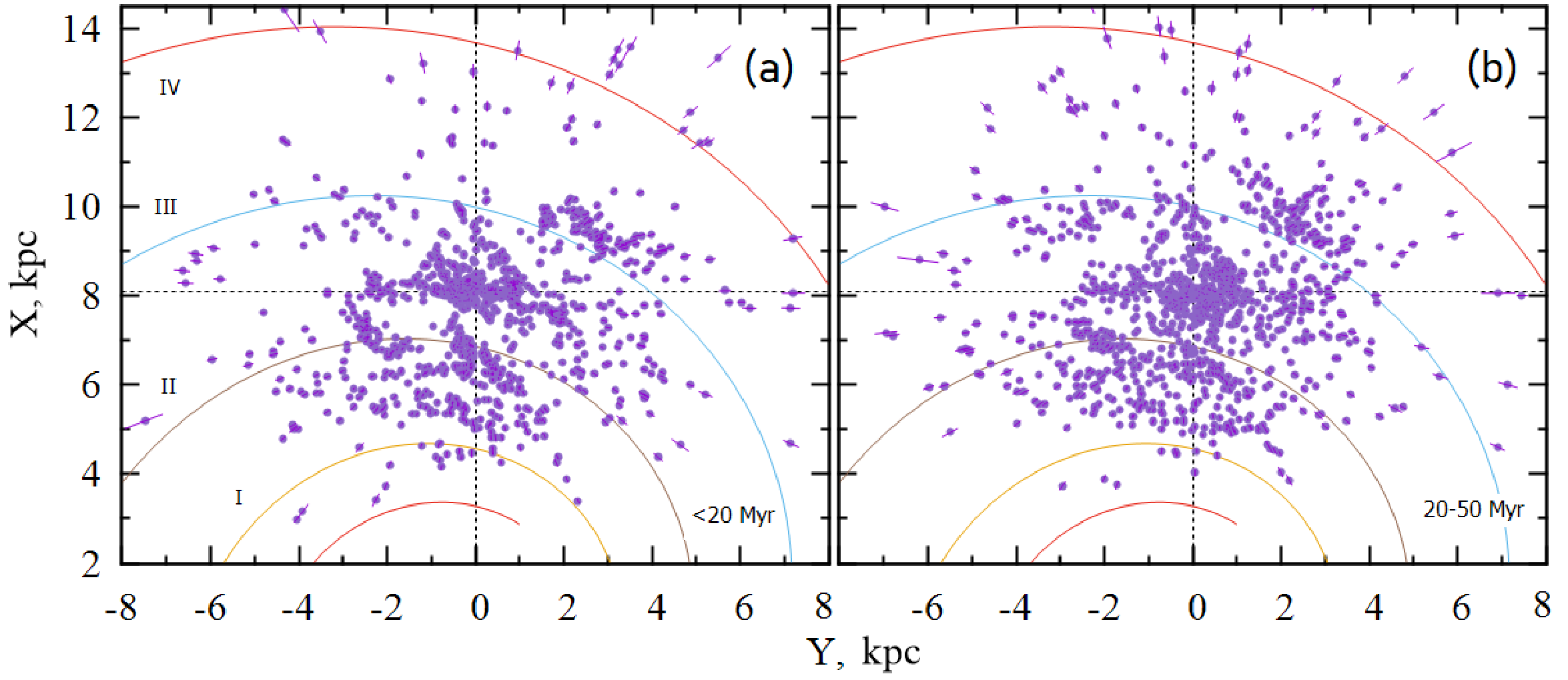}
  \caption{
Distribution of OSCs younger than 20~Myr (a) and with ages in the interval 20--50~Myr (b) in projection onto the Galactic $XY$ plane.}
 \label{f-XY}
\end{center}}
\end{figure}

Figure 1 gives the distribution of OSCs younger than 20~Myr and with ages in the interval 20--50 Myr in projection onto the Galactic $XY$ plane. The coordinate
system in which the $X$ axis is directed from the Galactic center to the Sun and the direction of the $Y$ axis coincides with the direction of Galactic
rotation is used. The four-armed spiral pattern with the pitch angle $i=-13^\circ$ (Bobylev and Bajkova 2014) constructed with $R_0=8.1$~kpc is shown; the Roman numerals number the following spiral arm segments: Scutum (I), Carina--Sagittarius (II), Perseus (III), and the Outer Arm (IV). It follows from the figure than in both cases the selected OSCs trace the spiral pattern fairly well.

 \section{METHODS}
From observations we have three stellar velocity components: the line-of-sight velocity $V_r$ and the two tangential velocity components $V_l=4.74r\mu_l\cos b$ and $V_b=4.74r\mu_b$ along the Galactic longitude $l$ and latitude $b,$ respectively, expressed in km s$^{-1}$. Here, 4.74 is the dimension coefficient and $r$ is
the stellar heliocentric distance in kpc. The proper motion components $\mu_l\cos b$ and $\mu_b$ are expressed in mas yr$^{-1}$. The velocities $U,V,W$ directed along the
rectangular Galactic coordinate axes are calculated via the components 
$V_r, V_l, V_b$:
 \begin{equation}
 \begin{array}{lll}
 U=V_r\cos l\cos b-V_l\sin l-V_b\cos l\sin b,\\
 V=V_r\sin l\cos b+V_l\cos l-V_b\sin l\sin b,\\
 W=V_r\sin b                +V_b\cos b,
 \label{UVW}
 \end{array}
 \end{equation}
where the velocity $U$ is directed from the Sun toward
the Galactic center, $V$ is in the direction of Galactic
rotation, and $W$ is directed to the north Galactic
pole. We can find two velocities, $V_R$ directed radially
away from the Galactic center and the velocity $V_{\rm circ}$
orthogonal to it pointing in the direction of Galactic
rotation from the following expressions:
 \begin{equation}
 \begin{array}{lll}
  V_{\rm circ}= U\sin \theta+(V_0+V)\cos \theta, \\
           V_R=-U\cos \theta+(V_0+V)\sin \theta,
 \label{VRVT}
 \end{array}
 \end{equation}
where the position angle $\theta$ obeys the relation $\tan\theta=y/(R_0-x)$, $x,y,z$ are the rectangular heliocentric coordinates of the star (the velocities $U,V,W$ are
directed along the corresponding $x,y,z$ axes), and $V_0$ is the linear rotation velocity of the Galaxy at the solar distance $R_0$. The velocities $V_R$ and $W$ are virtually independent of the pattern of the Galactic rotation curve. However, to analyze the periodicities in the tangential velocities, it is necessary to determine a smoothed Galactic rotation curve and to form the residual velocities $\Delta V_{\rm circ}$.

 \subsection{Basic Kinematic Equations}
To determine the parameters of the Galactic rotation curve, we use the equations derived from Bottlinger's formulas, in which the angular velocity $\Omega$ is
expanded into a series to terms of the second order of smallness in $r/R_0:$
\begin{equation}
 \begin{array}{lll}
 V_r=-U_\odot\cos b\cos l-V_\odot\cos b\sin l -W_\odot\sin b\\
 +R_0(R-R_0)\sin l\cos b\Omega^\prime_0
 +0.5R_0(R-R_0)^2\sin l\cos b\Omega^{\prime\prime}_0\\
      -f_R\cos\chi\cos(l+\theta)\cos b
 +f_\theta\sin\chi\sin(l+\theta)\cos b,
 \label{EQ-1}
 \end{array}
 \end{equation}
 \begin{equation}
 \begin{array}{lll}
 V_l= U_\odot\sin l-V_\odot\cos l-r\Omega_0\cos b\\
 +(R-R_0)(R_0\cos l-r\cos b)\Omega^\prime_0
 +0.5(R-R_0)^2(R_0\cos l-r\cos b)\Omega^{\prime\prime}_0\\
      +f_R\cos\chi\sin(l+\theta)
 +f_\theta\sin\chi\cos(l+\theta),
 \label{EQ-2}
 \end{array}
 \end{equation}
 \begin{equation}
 \begin{array}{lll}
 V_b=U_\odot\cos l\sin b + V_\odot\sin l \sin b -W_\odot\cos b\\
 -R_0(R-R_0)\sin l\sin b\Omega^\prime_0
  -0.5R_0(R-R_0)^2\sin l\sin b\Omega^{\prime\prime}_0\\
        +f_R\cos\chi\cos(l+\theta)\sin b
   -f_\theta\sin\chi\sin(l+\theta)\sin b,
 \label{EQ-3}
 \end{array}
 \end{equation}
where $R$ is the distance of the star from the Galactic rotation axis, 
$R^2=r^2\cos^2 b-2R_0 r\cos b\cos l+R^2_0.$ The velocities $(U,V,W)_\odot$ are the mean group velocity of the sample, are taken with the opposite sign, and
reflect the peculiar motion of the Sun; $\Omega_0$ is the angular
velocity of Galactic rotation at the solar distance $R_0,$ the parameters 
$\Omega^{\prime}_0$ and $\Omega^{\prime\prime}_0$are the corresponding
derivatives of the angular velocity, and $V_0=|R_0\Omega_0|$.

In this paper $R_0$ is taken to be $8.1\pm0.1$~kpc, according to the review by Bobylev and Bajkova (2021), where it was derived as a weighted mean of a large number of present-day individual estimates, and the four-armed model of the Galactic spiral pattern ($m=4$) is used.

According to the linear density wave theory (Lin and Shu 1964), the perturbations from the Galactic spiral density wave in the velocities $V_R$ and $\Delta V_{circ}$ are periodic and described by functions of the following form:
 \begin{equation}
 \begin{array}{lll}
                 V_R=f_R\cos\chi,\\
 \Delta V_{\rm circ}=f_\theta\sin\chi,
 \label{DelVRot}
 \end{array}
 \end{equation}
where $\chi$~is the radial phase of the spiral density wave,
 \begin{equation}
 \chi=m[\cot(i)\ln(R/R_0)-\theta]+\chi_\odot,
  \label{chi}
 \end{equation}
while $\chi_\odot$ is the radial phase of the Sun in the spiral density wave, $m$ is the number of spiral arms, $i$ is the pitch angle of the spiral pattern ($i<0$ for winding spirals), $f_R$ and $f_\theta$ are the amplitudes of the radial and
tangential velocity perturbations, respectively. The wavelength $\lambda$ (the distance between adjacent spiral arm segments measured along the radial direction) is
calculated from the relation
\begin{equation}
 2\pi R_0/\lambda=m\cot(|i|).
 \label{a-04}
\end{equation}
From the solution of the system of conditional equations (3)--(5), given (7), we can determine the following unknowns: $U_\odot,$ $V_\odot,$ $W_\odot,$ $\Omega_0,$ $\Omega^\prime_0,$ $\Omega^{\prime\prime}_0,$ $i,$ $\chi_\odot,$ $f_R$ and $f_\theta.$

 \subsection{Spectral Analysis}
Another approach is to find only six unknowns from the solution of the system of conditional equations (3)--(5) with the excluded parameters of the
spiral density wave: $U_\odot,$ $V_\odot,$ $W_\odot,$ $\Omega_0,$ $\Omega^\prime_0$ and $\Omega^{\prime\prime}_0.$ Then, the residual velocities $V_R,$ $\Delta V_{circ}$ and $W$ are formed. They are residual in the sense that (i) the peculiar solar motion with the values of $U_\odot,$ $V_\odot,$ $W_\odot$ found was subtracted from them and (ii) the Galactic rotation with the values of $\Omega_0,$ $\Omega^\prime_0$ and $\Omega^{\prime\prime}_0$ found was taken into account in the velocities 
$\Delta V_{circ}$.

Let there be a series of measured velocities $V_{R_n}$ (these can be the radial, $V_R,$ tangential, $\Delta V_{\rm circ},$ and vertical, $W,$ velocities), $n=1,\dots,N$, where$N$ is the number of objects. The objective of our spectral
analysis is to extract a periodicity from the data series
in accordance with the adopted model describing a
spiral density wave with parameters$f,$ $\lambda$~(or $i)$ and $\chi_\odot$/

Having taken into account the logarithmic behavior
of the spiral density wave and the position angles
of the objects $\theta_n$, our spectral (periodogram) analysis
of the series of velocity perturbations is reduced to
calculating the square of the amplitude (power spectrum)
of the standard Fourier transform (Bajkova and Bobylev 2012):
\begin{equation}
 \bar{V}_{\lambda_k} = \frac{1} {N}\sum_{n=1}^{N} V^{'}_n(R^{'}_n)
 \exp\biggl(-j\frac {2\pi R^{'}_n}{\lambda_k}\biggr),
 \label{29}
\end{equation}
where $\bar{V}_{\lambda_k}$ is the $k$th harmonic of the Fourier transform
with wavelength $\lambda_k=D/k$, $D$ is the period of the series being analyzed,
 \begin{equation}
 \begin{array}{lll}
 R^{'}_{n}=R_0\ln(R_n/R_0),\\
 V^{'}_n(R^{'}_n)=V_n(R^{'}_n)\times\exp(jm\theta_n).
 \label{21}
 \end{array}
\end{equation}
The sought-for wavelength $\lambda$ corresponds to the peak
value of the power spectrum $S_{\rm peak}$. The pitch angle
of the spiral density wave is found from Eq. (8). We
find the perturbation amplitude and phase by fitting
the harmonic with the wavelength found to the observational
data. The relation
 \begin{equation}
 f_R (f_\theta, f_W)=2\times\sqrt{S_{\rm peak}}.
 \label{Speak}
 \end{equation}
can also be used to estimate the perturbation amplitude.

 \subsection{Estimation of the Pattern Speed $\Omega_p$}
The relation (Rohlfs 1977) following from the linear
density wave theory by Lin and Shu (1964) underlies
the approach that we apply in this paper:
\begin{equation}
 \begin{array}{lll}
 \renewcommand{\arraystretch}{3.6}
 \displaystyle
\chi=m[\Omega_p-\Omega(R)] t+\ln\biggl(\frac{R}{R_0}\biggr)\cot i=
\varkappa\nu t+\ln\biggl(\frac{R}{R_0}\biggr)\cot i,
 \label{spiral-777}
 \end{array}
\end{equation}
where $\Omega=\Omega(R)$ is the angular velocity of Galactic
rotation, $\Omega_p$ is the spiral pattern speed, $\varkappa^2=4\Omega^2\left(1+\frac{\displaystyle R}{\displaystyle 2\Omega}
\frac{\displaystyle d\Omega}{\displaystyle dR}\right)$ is the epicyclic frequency ($\varkappa>0$), and $\nu=m(\Omega_p-\Omega)/\varkappa$ is the frequency with which a
test particle encounters the passing spiral perturbation.

 \subsubsection{Method I ($\nu$).}
On the other hand, $f_R$ and $f_\theta$ have the following form:
 \begin{equation}
 \renewcommand{\arraystretch}{2.0}
       f_R= {k A\over \varkappa}{\nu \over {1-\nu^2}} \Im^{(1)}_\nu(x),
 \label{Factors-1}
 \end{equation}
 \begin{equation}
  f_\theta= -{k A\over 2\Omega} {1 \over {1-\nu^2}} \Im^{(2)}_\nu(x),
 \label{Factors-2}
 \end{equation}
where $A$ is the amplitude of the spiral density wave
potential, $k=m \cot i/R$ is the radial wave number,
$\Im^{(1)}_\nu(x)$ and $\Im^{(2)}_\nu(x)$ are the reduction factors:
\begin{equation}
 \begin{array}{lll}
 \renewcommand{\arraystretch}{3.8}
 \displaystyle
  \Im^{(1)}_\nu(x)={{1-\nu^2}\over x} \biggl[1-{\nu\pi\over \sin(\nu\pi)}\times {1\over 2\pi}\int_{-\pi}^{+\pi} e^{-x(1+\cos(s))} \cos(\nu s) ds \biggr],
 \label{spiral-77}
 \end{array}
\end{equation}
\begin{equation}
 \begin{array}{lll}
 \displaystyle
  \Im^{(2)}_\nu(x)=(\nu^2-1) {\nu\pi\over \sin(\nu\pi)}
  \times{\partial\over\partial x}
  \Biggl[{1\over 2\pi}\int_{-\pi}^{+\pi} e^{-x(1+\cos(s))} \cos(\nu s) ds\Biggr],
 \label{spiral-99}
 \end{array}
\end{equation}
which are functions of the coordinate $x=k^2\sigma^2_R/\varkappa^2,$
where $\sigma_R$ is the root-mean-square value of the stellar
radial velocity dispersion. Relations (13)--(16) allow
 $\Omega_p$ to be determined after the substitution of the parameters
($f_R, f_\theta, \Omega_0, \Omega'_0,\sigma_R$) obtained from observations.

We estimate the amplitude of the spiral density wave potential $A$ based on the following well-known relation (Fern\'andez et al. 2008):
 \begin{equation}
 A=\frac{(R_0\Omega_0)^2 f_{r0} \tan i}{m},
 \label{f-r0}
 \end{equation}
where the ratio of the radial component of the gravitational force corresponding to the spiral arms to the total gravitational force of the Galaxy, $f_{r0},$ is taken to
be $0.04\pm0.01$ (Bajkova and Bobylev 2012).

 \subsubsection{Method II (Direct).}
The equation of a logarithmic spiral can be written as follows (Yuan 1969):
\begin{equation}
 \ln\biggl(\frac{R}{R_0}\biggr)=\tan i~\biggl(\theta+\frac{\chi-\chi_0}{m}-
 \Omega_p t\biggr),
 \label{spiral-44}
\end{equation}
where $t$ is the time elapsed since the birth of the object. A simple relation follows from Eq. (18):
\begin{equation}
  \Omega_p=\frac{\theta-\theta_{\rm birth}}{t},
 \label{Omega-Lin}
\end{equation}
where $\theta$ is the current position of the star, $\theta_{\rm birth}$ is the
position angle corresponding to the birthplace of the star, and $t$ is the age of the star. If the age $t$ in (10) is expressed in Myr, then multiplying the difference $\Delta \theta,$ expressed in radians by the coefficient 1023, we will
obtain $\Omega_p$ in km s$^{-1}$ kpc$^{-1}$.

Following Dias and L\'epine (2005), this $\Omega_p$ estimation method is called the direct one. It is applied in those cases where the space velocities of stars, their individual ages, and occasionally (e.g., Joshi and Malhotra 2022) their membership in a specific spiral arm are known. To determine the birthplaces of stars $\theta_{\rm birth}$, their Galactic orbits are integrated backward in time using an appropriate model of the Galactic gravitational potential.

In this paper we use a three-component axisymmetric model of the Galactic gravitational potential that contains the contributions from such subsystems
as the central bulge, the disk, and the dark matter halo. In this case, $\Phi(R,z)=\Phi_b(r(R,z))+\Phi_d(r(R,z))+\Phi_h(r(R,z))$, where $\Phi_b(r(R,z))$ is the contribution of the bulge, $\Phi_d(r(R,z))$ is the contribution
of the disk, and $\Phi_h(r(R,z))$ is the contribution of the dark matter halo. The Galactic bulge is modeled by a Plummer sphere (Plummer 1911):
 \begin{equation}
  \Phi_b(r)=-\frac{G M_b}{\sqrt{r^2+b_b^2}}.
  \label{bulge}
 \end{equation}
The disk is modeled by an oblate spheroid in the form proposed by Miyamoto and Nagai (1975):
 \begin{equation}
 \Phi_d(R,z)=-\frac{G M_d}{\sqrt{ R^2+\biggl[ a_d+\sqrt{z^2+b_d^2} ~\biggr]^2  }}.
 \label{disk}
\end{equation}
To represent the dark matter halo, we use the Navarro--Frenk--White model (Navarro et al. 1997):
 \begin{equation}
  \Phi_h(r)=-\frac{G M_h}{r} \ln {\biggl(1+\frac{r}{a_h}\biggr)},
 \label{halo-III}
 \end{equation}
where $M_b, M_d, M_h$ are the masses of the components, $b_b, a_d, b_d, a_h$ are the scale parameters of the components. In Bajkova and Bobylev (2016, 2017), the approach based on relations (21) and (22) corresponds to model III.

  \begin{table}[t]
  \caption[]{\small Estimates of the Galactic rotation parameters}
  \begin{center}  \label{Table-1} \small
  \begin{tabular}{|l|c|c|c|c|c|c|}\hline
 Parameters & $<50$~Myr &$<20$~Myr & $20-50$~Myr \\\hline
 $N_\star$                  &      2494    & 1332     &    1162 \\
 $N_\star$ with line-of-sight velocities &   1722  &  984     &     738 \\

 $U_\odot,$~km s$^{-1}$   & $8.43\pm0.24$ & $7.79\pm0.30$ & $9.22\pm0.36$ \\
 $V_\odot,$~km s$^{-1}$   & $8.70\pm0.27$ & $7.99\pm0.34$ & $9.65\pm0.41$ \\
 $W_\odot,$~km s$^{-1}$   & $7.88\pm0.22$ & $7.89\pm0.29$ & $7.85\pm0.34$ \\

 $\Omega_0,$~km s$^{-1}$ kpc$^{-1}$   &$ 29.18\pm0.13$  &$ 29.43\pm0.17$  & $ 29.00\pm0.18$  \\
 $\Omega^{'}_0,$~km s$^{-1}$ kpc$^{-2}$  &$-3.854\pm0.030$ &$-3.861\pm0.041$ & $-3.885\pm0.044$ \\
 $\Omega^{''}_0,$~km s$^{-1}$ kpc$^{-3}$ &$ 0.590\pm0.012$ &$ 0.592\pm0.016$ & $ 0.605\pm0.017$ \\

 $\sigma_R,$~km s$^{-1}$      &       13.55 &       12.54 &       14.66 \\
 $f_R,$~km s$^{-1}$           & $4.7\pm0.2$ & $5.5\pm0.4$ & $5.8\pm0.4$ \\
 $f_\theta,$~km s$^{-1}$      & $3.7\pm0.2$ & $4.2\pm0.4$ & $6.0\pm0.4$ \\
 \hline
 \end{tabular}\end{center}
 {\small $N_\star$~is the number of stars used; the estimates of $f_R$ and $f_\theta$ given here were found through our spectral analysis of the residual velocities $V_R$ and $\Delta V_{circ}$.}
 \end{table}
\begin{figure}[t]
{ \begin{center}
  \includegraphics[width=0.8\textwidth]{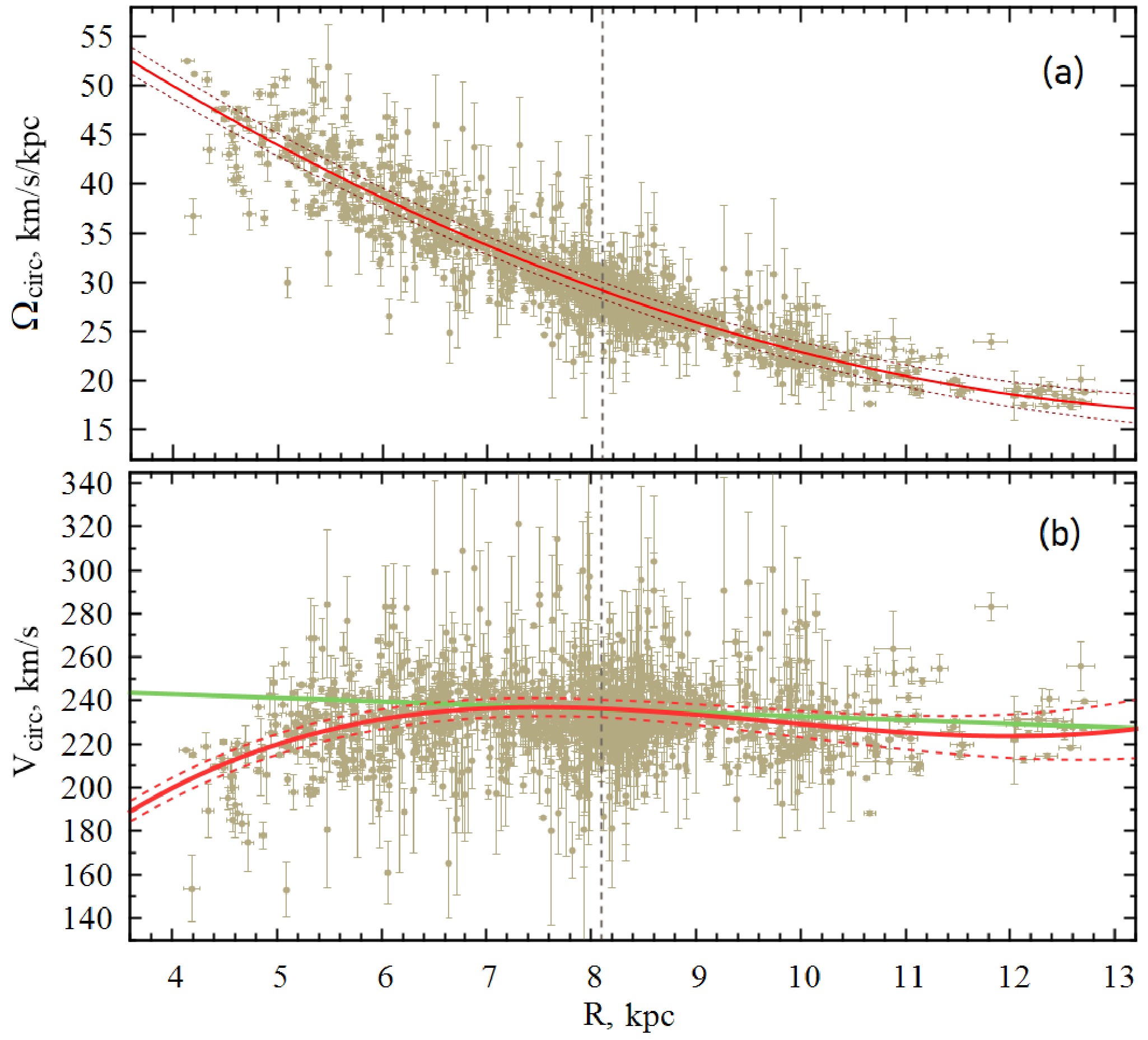}
\caption{
Angular, $\Omega_{\rm circ}$~(a), and linear, $V_{\rm circ}$~(b), circular rotation velocities of the sample of OSCs younger than 50 Myr versus
distance $R.$ Each panel gives the Galactic rotation curve found with an indication of the boundaries of the 1$\sigma$ confidence regions. The vertical dashed line marks the Sun's position; the green line on panel (b) gives the rotation curve from Eilers
et al. (2019).
}
 \label{f-rotation}
\end{center}}
\end{figure}

 \section{RESULTS}
The system of conditional equations (3)--(5) is solved by the least-squares methods (LSM) with weights of the form 
$w_r=S_0/\sqrt {S_0^2+\sigma^2_{V_r}},$
$w_l=S_0/\sqrt {S_0^2+\sigma^2_{V_l}}$ and
$w_b=S_0/\sqrt {S_0^2+\sigma^2_{V_b}},$ where $S_0$ is the ``cosmic'' dispersion and $\sigma_{V_r}, \sigma_{V_l}, \sigma_{V_b}$ are the dispersions of the corresponding observed velocities. $S_0$ is comparable to the root-mean-square residual
$\sigma_0$ in solving the conditional equations (3)--(5). In this paper we adopted $S_0=10$~km s$^{-1}$. The system of equations (3)--(5) was solved in several iterations using the $3\sigma$ criterion to eliminate the OSCs with large residuals.

 \subsection{Method I}
Table 1 gives our estimates of the Galactic rotation parameters found through the LSM solution of the system of conditional equations (3)--(5) with six
unknowns: $U_\odot,$ $V_\odot,$ $W_\odot,$ $\Omega_0,$ $\Omega^\prime_0$ and $\Omega^{\prime\prime}_0$.

For the three samples listed in the table we precalculated the corresponding dispersions of their velocities in the radial (along the $R$ axis) direction $\sigma_R$ that are required to properly apply this method.

\begin{figure}[t]
{ \begin{center}
  \includegraphics[width=0.85\textwidth]{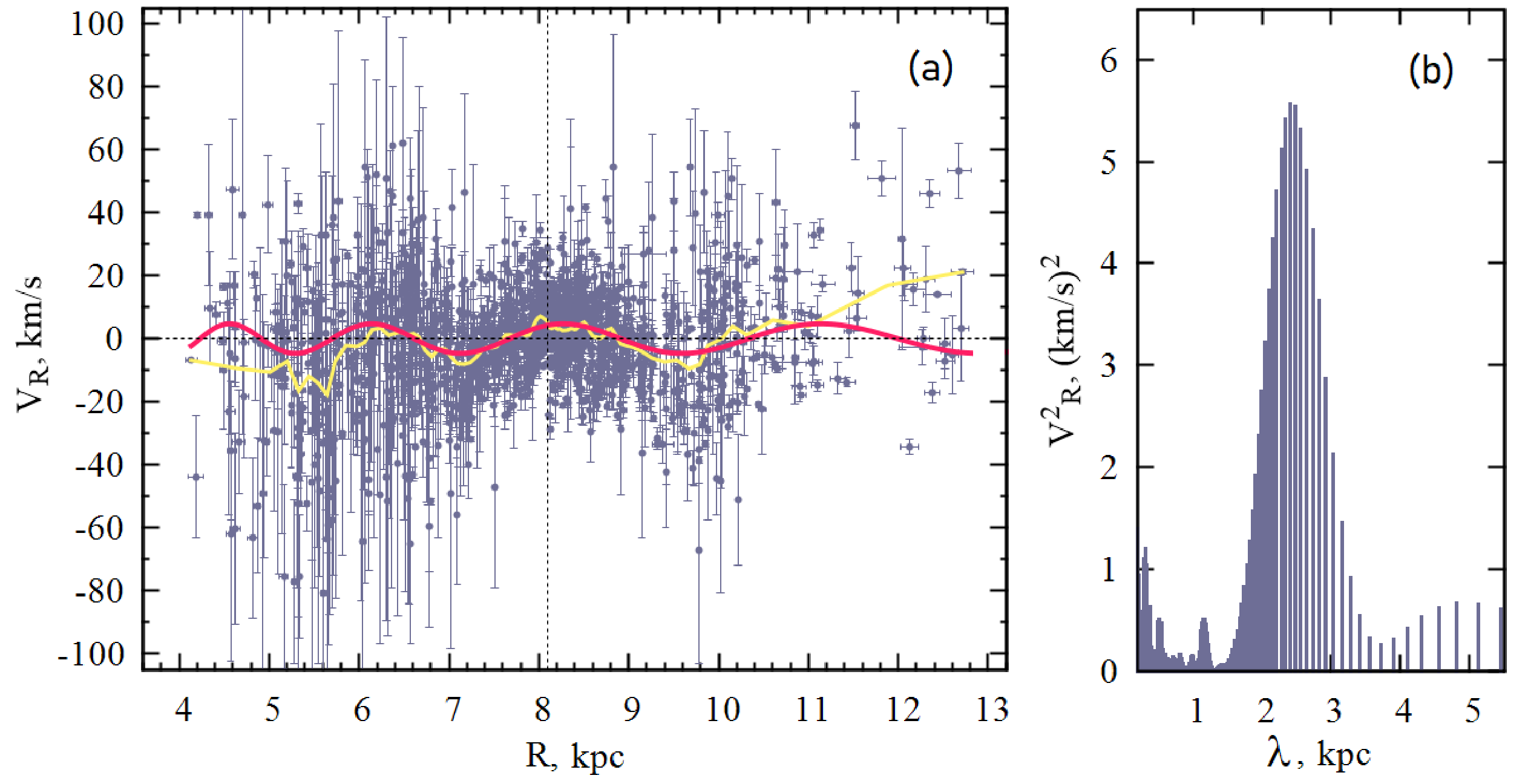}
  \caption{
Radial velocities $V_R$ versus distance $R$ for the sample of OSCs younger than 50~Myr (a) and their power spectrum(b). The averaged data are represented by the yellow line; the periodic curve found from our spectral analysis is indicated by the red line.}
 \label{f-Rad-LT-50}
\end{center}}
\end{figure}
\begin{figure}[t]
{ \begin{center}
  \includegraphics[width=0.85\textwidth]{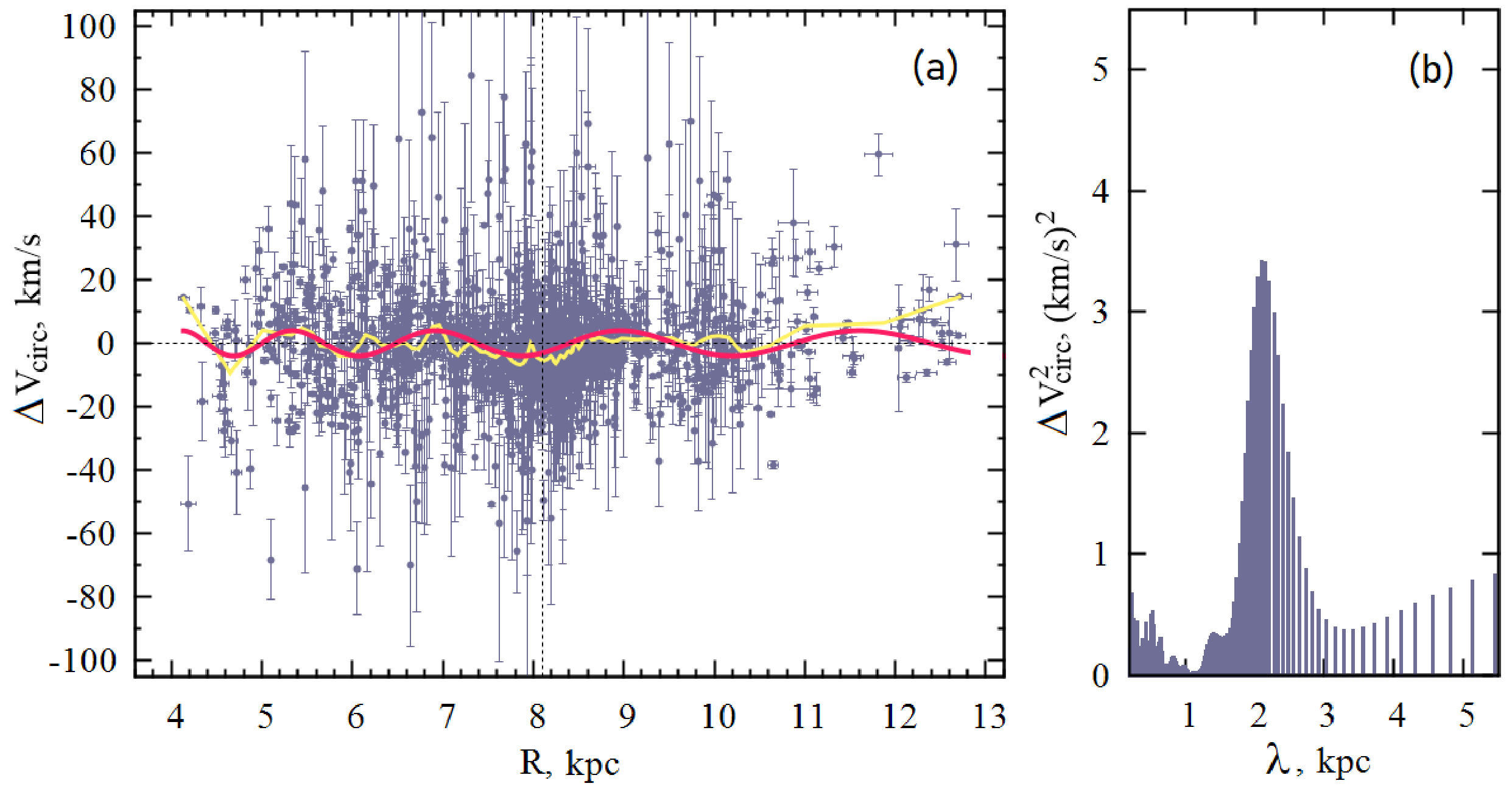}
  \caption{
Residual rotation velocities $\Delta V_{circ}$ versus distance $R$ for the sample of OSCs younger than 50~Myr (a) and their power spectrum (b). The averaged data are represented by the yellow line; the periodic curve found from our spectral analysis is indicated by the red line.}
 \label{f-Thet-LT-50}
\end{center}}
\end{figure}
\begin{figure}[t]
{ \begin{center}
  \includegraphics[width=0.85\textwidth]{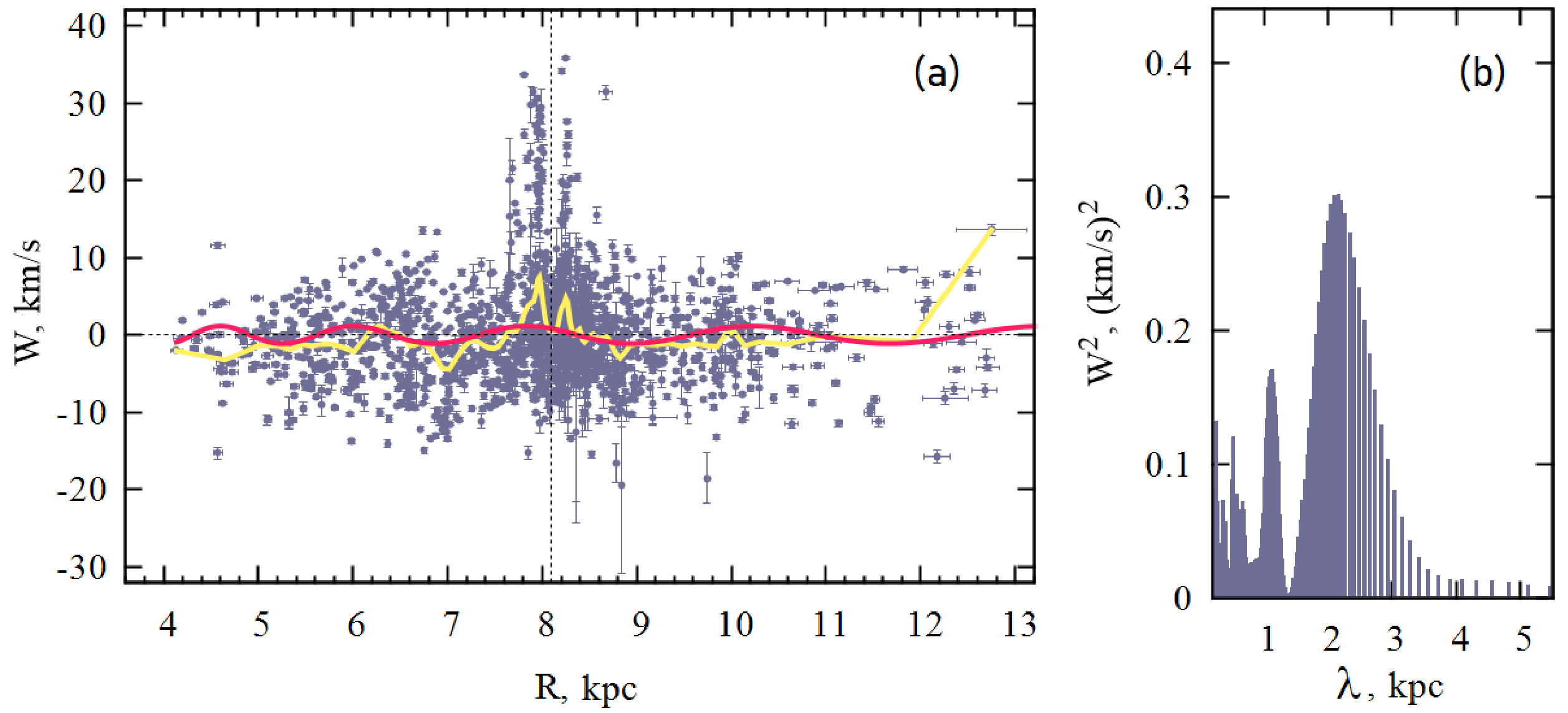}
  \caption{
Velocities $W$ versus distance $R$ for the sample of OSCs younger than 50~Myr (a) and their power spectrum(b). The averaged data are represented by the yellow line; the periodic curve found from our spectral analysis is indicated by the red line.
}
 \label{f-W-50}
\end{center}}
\end{figure}

The velocity perturbations $f_R$ and $f_\theta$ given in the table were estimated through our spectral analysis of the residual velocities $V_R$ and $\Delta V_{circ}.$ In particular, from the sample of OSCs younger than 50 Myr we also obtained the following estimates: $\lambda_R=2.4\pm0.2$~kpc and $\lambda_\theta=2.1\pm0.2$~kpc, $(\chi_\odot)_R=-154\pm11^\circ$ and $(\chi_\odot)_\theta=-129\pm10^\circ$. Our spectral analysis of the vertical velocities of this sample showed the presence
of a very strong spike within about 0.5 kpc of the Sun. The periodic perturbations in the vertical OSC velocities associated with the spiral density wave are also present but have a very small amplitude. For example, from the vertical velocities we obtained the following estimates: $f_W=1.1\pm0.4$~km s$^{-1}$, $\lambda_W=2.2\pm0.2$~kpc and $(\chi_\odot)_W=-135\pm12^\circ$. The results of our spectral analysis for the sample of OSCs younger than 50~Myr are displayed in Figs. 2--5.

Applying the method described in Subsection 2.3.1
to the sample of OSCs younger than 50 Myr yielded
the following value of the spiral pattern speed:
 \begin{equation}
 \Omega_p=24.26\pm0.52~\hbox{km s$^{-1}$ kpc$^{-1}$}.
 \label{Om-p-I}
\end{equation}
The corotation radius can be calculated from the relation
derived by equating the linear Galactic rotation
velocity and the rotation velocity of the spiral pattern
found:
\begin{equation}
R_{\rm cor}=R_0+(\Omega_p-\Omega_0)/\Omega'_0 .
 \label{R-cor}
\end{equation}
With $\Omega_0$ and $\Omega'_0$ taken from the first column of Table 1
we find
\begin{equation}
 \begin{array}{lll}
R_{\rm cor}=9.38\pm0.13~\hbox{kpc}=(1.16\pm0.02)R_0.
 \label{Rcor-I}
 \end{array}
\end{equation}

 \subsection{Method~II}
Here, to estimate $\Omega_p$, we apply an approach similar to the first method (Subsection 2.3.1). The difference is that we estimate the velocity perturbations 
$f_R$ and $f_\theta$ through the LSM solution of the system of conditional equations (3)--(5). For example, from the sample of OSCs younger than 50 Myr we obtained the following estimates:
 \begin{equation}
 \label{solution-I}
 \begin{array}{lll}
  U_\odot=7.95\pm0.26~\hbox{km s$^{-1}$},\\
  V_\odot=9.85\pm0.28~\hbox{km s$^{-1}$},\\
  W_\odot=7.39\pm0.22~\hbox{km s$^{-1}$},\\
      \Omega_0 = 27.46\pm0.15~\hbox{km s$^{-1}$ kpc$^{-1}$},\\
  \Omega^{'}_0 =-3.800\pm0.028~\hbox{km s$^{-1}$ kpc$^{-2}$},\\
 \Omega^{''}_0 =~0.640\pm0.014~\hbox{km s$^{-1}$ kpc$^{-3}$},\\
         i =-11.8\pm0.7^\circ,\\
  \chi_\odot= -143\pm4^\circ,\\
       f_R =-3.32\pm0.22~\hbox{km s$^{-1}$},\\
  f_\theta =-2.48\pm0.16~\hbox{km s$^{-1}$}
 \end{array}
 \end{equation}
at fixed $R_0=8.1\pm0.1.$ As a result of implementing
the algorithm described in Subsection 2.3.1, we obtained
 \begin{equation}
 \Omega_p=23.45\pm0.53~\hbox{km s$^{-1}$ kpc$^{-1}$}.
 \label{Om-p-II}
\end{equation}
With $\Omega_0$ and $\Omega'_0$ taken from the solution (26) we find
\begin{equation}
 \begin{array}{lll}
R_{\rm cor}=9.16\pm0.14~\hbox{kpc}=(1.13\pm0.02)R_0.
 \end{array}
 \label{Rcor-II}
\end{equation}

 \subsection{Method~III}
Here, we present the results of applying the direct method described in Subsection 2.3.2. Based on the integration of the Galactic orbits for OSCs younger
than 50~Myr, we found
 \begin{equation}
 \Omega_p=28.9\pm2.8~\hbox{km s$^{-1}$ kpc$^{-1}$}.
 \label{Om-p-III}
\end{equation}
To reduce the errors in the velocities of distant OSCs, we used a sample of $\sim$1500 OSCs located within 3~kpc of the Sun.

With $\Omega_0$ and $\Omega'_0$ taken from the first column of
Table 1 we find
\begin{equation}
 \begin{array}{lll}
 R_{\rm cor}=8.17\pm0.73~\hbox{kpc}=(1.01\pm0.09)R_0.
 \end{array}
 \label{Rcor-III}
\end{equation}

 \section{DISCUSSION}
In Bobylev and Bajkova (2022) we analyzed a sample of $\sim$150 masers with measured VLBI trigonometric parallaxes. We used masers with relative trigonometric parallax errors less than 10\%. From them we found the group velocity components
($(U,V,W)_\odot=(9.15,12.81,8.93)\pm(0.86,0.86,0.75)$~km s$^{-1}$ and the following parameters of the angular velocity of Galactic rotation:  
$\Omega_0=30.18\pm0.38$~km s$^{-1}$ kpc$^{-1}$,
$\Omega^{'}_0=-4.368\pm0.077$~km s$^{-1}$ kpc$^{-2}$ and
$\Omega^{''}_0=0.845\pm0.037$~km s$^{-1}$ kpc$^{-3}$, where the
linear rotation velocity of the Galaxy at the solar
distance $R_0$ was $244.4\pm4.3$~km s$^{-1}$ (for the adopted $R_0=8.1\pm0.1$~kpc).

As can be seen from our comparison with the results presented in Table 1, due to the large number of young OSCs and the first--class measurements,
the Galactic rotation parameters are determined from
them with small errors. In particular, the circular rotation velocity of the solar neighborhood around the Galactic center found from the sample of OSCs
younger than 50 Myr is $V_0=236.4\pm3.1$~km s$^{-1}$ (for $R_0=8.1\pm0.1$~kpc), while the constructed Galactic rotation curve is in excellent agreement with the
present-day results of other authors (Fig. 2b).

The parameters of the Galactic spiral density wave $f_R$ and $f_\theta$ found in this paper by two methods are in good agreement with the estimates obtained by
various authors from the most important spiral structure
tracers. An overview of such estimates can be found, for example, in Bobylev and Bajkova (2022). In particular, from data on 134 masers Bobylev and
Bajkova (2022) found $f_R=8.1\pm1.4$~km s$^{-1}$, $f_\theta=6.1\pm1.7$~km s$^{-1}$ and $f_W=5.2\pm1.5$~km s$^{-1}$.

Note that the wave in the vertical velocities $W$
with a small amplitude fW is clearly visible only in
the velocities of the youngest objects, such as masers
(Bobylev and Bajkova 2015; Rastorguev et al. 2017).
We see a low-amplitude wave in Fig. 5. In our
opinion, this suggests a high quality of the OSC
measurements. At the same time, in Fig. 5 we can
see an unusual spike of positive vertical velocities in a
relatively small solar neighborhood. This interesting
effect, possibly associated with to the structure of the
Gould Belt, requires a separate study.

In Bobylev and Bajkova (2023) we analyzed three stellar samples --- Galactic masers with measured trigonometric parallaxes, OB2 stars, and Cepheids.
The following estimates of $\Omega_p$ were obtained from
these data: $24.61\pm2.06$~km s$^{-1}$ kpc$^{-1}$, $24.71\pm1.29$~km s$^{-1}$ kpc$^{-1}$ and $25.98\pm1.37$~km s$^{-1}$ kpc$^{-1}$ from the samples of masers,
OB2 stars, and Cepheids, respectively. The values of the corotation radius 
$R_{\rm cor}/R_0$ for these three samples were $1.16\pm0.09$, $1.15\pm0.06$ and $1.09\pm0.06$.

In Bobylev and Bajkova (2023) we also gave a table with 17 estimates of $\Omega_p$ obtained by various authors using the three methods from young objects.
The estimates of $\Omega_p$ obtained by various authors in
recent years were shown to lie in a fairly wide range of
$\Omega_p$: 18--32~km s$^{-1}$ kpc$^{-1}$.

Note that the values of $\Omega_p$ that we obtained in this paper by methods I and II from three samples are in excellent agreement both between themselves and
with the estimates from Bobylev and Bajkova (2023).

The estimate of (29) was obtained by method III with a greater uncertainty. At the same time, this estimate is consistent with the results of applying
method III by other authors, as can be seen from Table 2 in Bobylev and Bajkova (2023). Note, for example, the result of applying method III to the
analysis of OSCs in Dias et al. (2019): $\Omega_p=28.2\pm2.1$~km s$^{-1}$ kpc$^{-1}$.

 \section*{CONCLUSIONS}
Young OSCs from the catalogue by Hunt and Reffert (2023) served us as the basis for our kinematic analysis. Their mean proper motions, line-of-sight
velocities, and distances were calculated by these
authors based on data from the Gaia~DR3 catalogue.
An important advantage of the catalogue by Hunt and Reffert (2023) is the availability of line-of-sight velocities for a high percentage of OSCs, which allowed a full-fledged analysis of the space velocities of these clusters to be performed.

First, using 2494 OSCs younger than 50 Myr, we estimated the Galactic rotation parameters. In particular, we found the solar velocity components
$(U,V,W)_\odot=(8.43,8.70,7.88)\pm(0.24,0.27,0.22)$~km s$^{-1}$ and the following terms of the expansion of the angular velocity of Galactic rotation:
 $\Omega_0 =29.18\pm0.12$~km s$^{-1}$ kpc$^{-1}$,
 $\Omega^{'}_0=-3.854\pm0.030$~km s$^{-1}$ kpc$^{-2}$ and
 $\Omega^{''}_0=0.590\pm0.012$~km s$^{-1}$ kpc$^{-3}$.
Here, the circular rotation velocity of the solar neighborhood around the Galactic
center is $V_0=236.4\pm3.1$~km s$^{-1}$ for the adopted distance $R_0=8.1\pm0.1$~kpc.

To estimate $\Omega_p$, we applied three methods based on the linear Lin--Shu spiral density wave theory. Based on the first method, which is most reliable in our view,
and using the velocity perturbations $f_R$ and $f_\theta,$ found through our spectral analysis of the radial, $V_R,$ and residual rotation, $\Delta V_{\rm circ},$ velocities, we obtained an estimate of $\Omega_p=24.26\pm0.52$~km s$^{-1}$ kpc$^{-1}$. Here, the velocity perturbations $f_R=4.72\pm0.24$~km s$^{-1}$
and $f_\theta=3.69\pm0.20$~km s$^{-1}$ were estimated from the
space velocities of 1722 OSCs younger than 50 Myr.

In the second method $f_R=3.32\pm0.22$~km s$^{-1}$ and $f_\theta=2.48\pm0.16$~km s$^{-1}$ were found by solving the basic kinematic equations together with the
Galactic rotation parameters. Here, we obtained an
estimate of $\Omega_p=23.45\pm0.53$~km s$^{-1}$ kpc$^{-1}$.

Using the third method, which is based on our analysis of the position angles for 2494 OSCs younger than 50 Myr at their birth time $\theta_{\rm birth}$, we found
$\Omega_p = 28.9\pm2.8$~km s$^{-1}$ kpc$^{-1}$.

We noticed an unusual spike of positive vertical velocities $W$ in the local solar neighborhood with a radius $\sim$0.5 kpc. This effect, possibly associated not
with the spiral structure but with peculiarities of the Gould Belt, requires a separate study.

 \subsubsection*{ACKNOWLEDGMENTS}
We are grateful to the referees for their useful remarks that contributed to an improvement of the paper.

 \subsubsection*{REFERENCES}
 \small

\quad~~1. L. H. Amaral and J. R. D. L\'epine, Mon. Not. R. Astron. Soc. 286, 885 (1997).

2. A. T. Bajkova and V. V. Bobylev, Astron. Lett. 38, 549 (2012).

3. A. T. Bajkova and V. V. Bobylev, Astron. Lett. 42, 567 (2016).

4. A. T. Bajkova and V. V. Bobylev, Open Astron. 26, 72 (2017).

5. V. V. Bobylev, A. T. Bajkova, and A. S. Stepanishchev, Astron. Lett. 34, 515 (2008).

6. V. V. Bobylev and A. T. Bajkova, Mon. Not. R. Astron. Soc. 437, 1549 (2014).

7. V. V. Bobylev and A. T. Bajkova, Mon. Not. R. Astron. Soc. 447, L50 (2015).

8. V. V. Bobylev and A. T. Bajkova, Astron. Lett. 45, 109 (2019).

9. V. V. Bobylev and A. T. Bajkova, Astron. Rep. 65, 498 (2021).

10. V. V. Bobylev and A. T. Bajkova, Astron. Lett. 48, 376 (2022).

11. V. V. Bobylev and A. T. Bajkova, Astron. Lett. 49, 110 (2023).

12. A. G. A. Brown, A. Vallenari, T. Prusti, J. H. J. de Bruijne, C. Babusiaux, C. A. L. Bailer-Jones, M. Biermann, D. W. Evans, et al. (Gaia Collab.), Astron.
Astrophys. 616, 1 (2018).

13. A. G. A. Brown, A. Vallenari, T. Prusti, J. H. J. de Bruijne, C. Babusiaux, M. Biermann, O.L. Creevely, D.W. Evans, et al. (Gaia Collab.), Astron. Astrophys.
649, 1 (2021).

14. D. Camargo, C. Bonatto, and E. Bica, Mon. Not. R. Astron. Soc. 450, 4150 (2015).

15. T. Cantat-Gaudin, F. Anders, A. Castro-Ginard, C. Jordi, M. Romero-Gomez, C. Soubiran, L. Casamiquela, Y. Tarricq, et al., Astron. Astrophys. 640, A1 (2020).

16. W. S. Dias, J. R. D. L\'epine, and B. S. Alessi, Astron. Astrophys. 376, 44 (2001).

17. W. S. Dias and J. R. D. L\'epine, Astrophys. J. 629, 825 (2005).

18. W. S. Dias, M. Assafin, V. Fl\'orio, B. S. Alessi, and V. Libero, Astron. Astrophys. 446, 949 (2006).

19. W. S. Dias, H. Monteiro, J. R. D. L\'epine, and D. A. Barros, Mon. Not. R. Astron. Soc. 486, 5726 (2019).

20. W. S. Dias, H. Monteiro, A. Moitinho, J. R. D. Lepine, G. Carraro, E. Paunzen, B. Alessi, and L. Villela, Mon. Not. R. Astron. Soc. 504, 356 (2021).

21. A.-C. Eilers, D. W. Hogg, H.-W. Rix, and M. K. Ness, Astrophys. J. 871, 120 (2019).

22. D. Fern\'andez, F. Figueras, and J. Torra, Astron. Astrophys. 480, 735 (2008).

23. E. V. Glushkova, A. K. Dambis, A. M. Mel'nik, and A. S. Rastorguev, Astron. Astrophys. 329, 514 (1998).

24. C. J. Hao, Y. Xu, L. G. Hou, S. B. Bian, J. J. Li, Z. Y. Wu, Z. H. He, Y. J. Li, and D. J. Liu, Astron. Astrophys. 652, A102 (2021).

25. C. J. Hao, Y. Xu, Z. Y. Wu, Z. H. Lin, D. J. Liu, and
Y. J. Li, Astron. Astrophys. 660, A4 (2022).

26. E. L. Hunt and S. Reffert, Astron. Astrophys. 673, A114 (2023).

27. Y. C. Joshi and S. Malhotra, arXiv: 2212.09384 (2022).

28. T. C. Junqueira, C. Chiappini, J. R. D. L\'epine, I. Minchev, and B. X. Santiago, Mon. Not. R. Astron. Soc. 449, 2336 (2015).

29. M. A. Kuhn, L. A. Hillenbrand, A. Sills, E. D. Feigelson,
and K. V. Getman, Astrophys. J. 870, 32 (2019).

30. J. R. D. L\'epine, W. S. Dias, and Yu. Mishurov, Mon.
Not. R. Astron. Soc. 386, 2081 (2008).

31. C. C. Lin and F. H. Shu, Astrophys. J. 140, 646 (1964).

32. A. V. Loktin and M. E. Popova, Astrophys. Bull. 74, 270 (2019).

33. M. Miyamoto and R. Nagai, Publ. Astron. Soc. Jpn. 27, 533 (1975).

34. H. Monteiro, D. A. Barros, W. S. Dias, and J. R. D. L\'epine, Front. Astron. Space. Sci. 8, 62 (2021).

35. S. Naoz and N. J. Shaviv, New Astron. 12, 410 (2007).

36. J. F. Navarro, C. S. Frenk, and S. D. M. White, Astrophys. J. 490, 493 (1997).

37. A. E. Piskunov, N. V. Kharchenko, S. R\"oser, E. Schilbach, and R.-D. Scholz, Astron. Astrophys. 445, 545 (2006).

38. H. C. Plummer, Mon. Not. R. Astron. Soc. 71, 460 (1911).

39. M. E. Popova and A. V. Loktin, Astron. Lett. 31, 663 (2005).

40. M. E. Popova, Astrophys. Bull. 78, 134 (2023).

41. T. Prusti, J. H. J. de Bruijne, A. G. A. Brown, A. Vallenari, C. Babusiaux, C. A. L. Bailer-Jones, U. Bastian, M. Biermann, et al. (Gaia Collab.), Astron.
Astrophys. 595, A1 (2016).

42. A. S. Rastorguev, M. V. Zabolotskikh, A. K. Dambis, N. D. Utkin, V. V. Bobylev, and A. T. Baikova, Astrophys.mBull. 72, 122 (2017).

43. K. Rohlfs, {\it Lectures on Density Wave Theory} (Springer, Berlin, 1977).

44. Y. Tarricq, C. Soubiran, L. Casamiquela, T. Cantat-Gaudin, L. Chemin, F. Anders, T. Antoja, M. Romero-Gomez, et al., Astron. Astrophys. 647, A19 (2021).

45. A. Vallenari, A. G. A. Brown, T. Prusti, J. H. J. de Bruijne, F. Arenou, et al. (Gaia Collab.), arXiv: 2208.0021 (2022).

46. C. Yuan, Astrophys. J. 158, 889 (1969).

47. M. V. Zabolotskikh, A. S. Rastorguev, and A. K. Dambis, Astron. Lett. 28, 454 (2002).

 \end{document}